\documentclass[10pt,journal,a4paper,twoside]{./EWTEC}
\usepackage{cite}
\usepackage[pdftex]{graphicx}
\usepackage{textcomp}
\usepackage{amsmath}
\usepackage{booktabs}
\usepackage{threeparttable}
\usepackage[caption=false,font=footnotesize]{subfig}
\usepackage{url}

\begin{document}

\title{Ocean wave transmission, reflection and absorption by rows of vertical structures along the coastline}

\author{Alexis M{\'e}rigaud,
        Benjamin Thiria,
        and~Ramiro Godoy-Diana
\thanks{In \emph{Proceedings of the 11th European Wave and Tidal Energy Conference} 5-9th Sept 2021, Plymouth, UK.  
This project has received funding from the European Union's Horizon 2020 research and innovation programme under the Marie Sk\l{}odowska-Curie grant agreement No 842967. }
\thanks{A. M{\'e}rigaud, B. Thiria and R. Godoy-Diana are with the Laboratoire de Physique et M{\'e}canique des Milieux H{\'e}t{\'e}rog{\'e}nes (PMMH), CNRS UMR 7636, ESPCI Paris---PSL University, Sorbonne Universit{\'e}, Universit{\'e} de Paris, 75005 Paris, France. Contact e-mail: alexis.merigaud@espci.fr}
}

\markboth{}{M{\'e}rigaud \MakeLowercase{\textit{et al.}}: Ocean wave transmission, reflection and absorption by rows of vertical structures along the coastline}

\maketitle

\begin{abstract}
Large arrays of wave-absorbing structures could serve the double objective of  coastal protection against erosion and clean, renewable electrical power production.
In this work, the principle of an artificial canopy is explored, which consists of vertical structures, arranged in rows parallel to the coastline. Sea waves, which propagate towards the shore, interact with the obstacle rows. A part of the wave energy is reflected back towards the ocean, another part is transmitted to the shoreline, while the rest of the energy is, in theory, available for energy production (although losses, due to viscous effects within the fluid, or imperfect efficiency of the power conversion mechanism, will unavoidably take place).
First, a simple geometric representation of the reflection/transmission properties of individual, fixed rows is presented. In the case of moving rows, relationships are drawn between the internal stiffness and damping parameters of the devices, on the one hand, and their reflection, transmission and absorption characteristics, on the other hand.
Array properties are then examined, depending on both individual row design parameters and row-to-row spacing values, using the wide-spacing approximation. A numerical case study illustrates the capabilities of the proposed modelling framework, with arrays of vertical, oscillating rectangular plates. The transmitted, reflected and absorbed wave spectra are examined, along with their dependencies on individual oscillator control tuning and array design parameters.
\end{abstract}

\begin{EWTECkeywords}
Oscillating wave surge converter, WEC arrays, optimal control, wave spectrum, Bragg resonances
\end{EWTECkeywords}

%
\EWTECpeerreviewmaketitle

\section{Introduction}

\EWTECPARstart{O}{cean} waves transport vast amounts of energy, with the potential to be transformed into useful, exploitable energy forms \cite{Callaway:2007}.  For instance, in Europe, wave energy could be a significant contributor to the electricity supply, with an estimated 300 to 400 GW potential along European Atlantic coastlines alone \cite{Babarit_2017_Book}. In addition to purveying clean, renewable power, large arrays of wave energy absorbing structures may also serve the objective of mitigating coastal erosion along the shoreline \cite{Nove_2018}, by mimicking the wave reduction effect of natural coastal defences such as mangroves, salt-marshes or seagrass and kelp beds \cite{Abanades:2015,Narayan:2016}. Such coastal protection effects are only seldom addressed in the wave energy literature; see however \cite{Clemente_2021_review} for a recent review of possible synergies between wave energy harvesting and other applications.

Among wave energy converter concepts, oscillating wave surge converters (OWSCs), which primarily exploit the horizontal fluid motion, are identified as one of the most promising and mature technological options \cite{Babarit_2017_Book}. Traditional OWSC designs consist of a rigid flap with a flat geometry, pitching around a rotation point, which can be fixed to the sea bed or floating \cite{Sarkar:2014}. Recent work has investigated the benefits of an array of rigid flaps with curved geometry \cite{Michele_2019}. Alternatively, \cite{Nove_2018,Nove2018_PhD} has explored an artificial canopy of bio-inspired, flexible OWSCs, with the two objectives of coastline protection and energy harvesting. These arrays of flexible structures may indeed present benefits in terms of survivability and absorption capabilities, in the same way that aquatic vegetation can withstand and dissipate surface wave energy \cite{Koehl:1977,Denny:2002}. 

From the theoretical point of view, on the one hand, each element of the artificial canopy acts as an oscillator, with its intrinsic natural frequency and damping coefficient - that is the most commonly-adopted perspective in the wave energy literature. On the other hand, the array as a whole can be viewed as a metamaterial, with properties related to its internal structure, such as the existence of crystallographic effects analog to those observed in solid-state physics or acoustics---e.g. Bragg resonances \cite{Garnaud:2009,Rey:2018,Arnaud:2017}. In a broader context than wave energy harvesting, several studies have explored those ideas in different systems, designed with a view to  water wave propagation engineering, ranging from the refraction phenomena of water waves propagating through an array of bottom-mounted structures \cite{Hu:2005,Arnaud:2017}, to the tuning of the sea bed topography \cite{Davies:1984,Berraquero:2013}, or the deployment of floating membranes with a crystalline array of defects that confer unique propagation features to the resulting hydroelastic waves \cite{Domino:2020}.

In this work, an artificial canopy of generic OWSCs is numerically studied, where the devices are arranged into a number of rows parallel to the coastline. Each row is considered infinite, and consists of regularly-spaced devices. Incident waves can be reasonably assumed long-crested, with crests also parallel to the coastline, due to nearshore refraction effects. The OWSC array behaviour is mathematically and numerically determined using a recursive approach, based on the so-called \textit{wide-spacing approximation} (WSA) \cite{Evans_1990b}, more often termed ``plane wave approximation'' in the wave energy literature \cite{folley2012review}. The principle of the approach employed here is as follows: First, solving the hydrodynamic diffraction-radiation problem for any row, taken in isolation, allows the determination of complex wave transmission and reflection coefficients for that specific row. Then assuming that transmission and reflection coefficients have been calculated for every such row, and relying on the WSA, which neglects evanescent wave modes in the calculation of interaction effects, the reflection and transmission properties of the whole array can be solved recursively, as new rows are successively added to the array. Furthermore, in the special case of regularly-spaced, identical rows, analytical formulae can be derived for the global transmission and reflection coefficients.

The WSA, on which the recursion outlined above relies, is rarely used in the wave energy literature (see \cite{folley2012review} and references therein), and encounters theoretical limitations which must not be overlooked. In particular, the row-to-row spacing should be, in theory, larger than the wavelength, whence the terminology \textit{wide-spacing approximation}, while individual device dimensions must be assumed smaller than the wavelength. Nevertheless, a number of numerical and experimental results, e.g. \cite{McIver:1993}, Chap. 3 of \cite{Li_2006}, Chap. 6 of \cite{Linton:2001},  suggest that the WSA remains accurate, even when the former underlying assumption is clearly violated.
Experimental results in a small scale wave tank, recently reported in \cite{Merigaud_2021}, confirm the appropriateness of the WSA to model interaction effects in an array of two and three rows of flexible OWSCs, even when the row-to-row spacing is only a fraction of the wavelength. Overall, for the present problem, the proposed WSA approach seems appropriate and computationally attractive, because of the large array size to be treated. In spite of its relative simplicity, the approach highlights a rich variety of array behaviours, and allows the determination of quantities of interests, in particular the reflected and transmitted wave fields, as well as the array energy absorption.

This study begins, in Section \ref{s:singlerow}, by showing the fundamental connections that exist between, on the one hand, the distinctive features of a single OWSC row, i.e. its geometrical, hydromechanical and power take-off (PTO) characteristics, and, on the other hand, the corresponding transmission, reflection and energy absorption properties. The aforementioned connections are visually represented in a simple geometrical framework. Section \ref{s:wsa} briefly explains how, based on the WSA, the transmission, reflection and absorption properties of an array with multiple rows can be recursively determined. An analytical solution is given for the case of an infinite number of identical, regularly-spaced rows. A numerical case study is described in Section \ref{s:casestudy}, and its results are reported in Section \ref{s:results}. The case of fixed structures is first examined, before arrays of moving structures, in other words, OWSCs, are investigated. The reflected, transmitted and absorbed wave power spectra are given special attention. Finally, in Section \ref{s:conclusion}, results are summarised and discussed, and avenues for future work are outlined.

\section{Reflection, transmission and absorption by a periodic row of oscillating wave surge converters}
\label{s:singlerow}

In this section, the properties of a single row of obstacles, in terms of wave transmission, reflection and absorption, are examined. We follow \cite{Merigaud_2021} in \ref{ss:problem} and \ref{ss:fixedstructure} for the presentation of the classical theoretical framework, while \ref{ss:movingstructures} introduces the case of moving, wave-absorbing structures and the effect of their control parameters onto their transmission, reflection and absorption properties.

\subsection{Problem formulation and notations}
\label{ss:problem}

\begin{figure}
    \centering
    \includegraphics[width = \columnwidth]{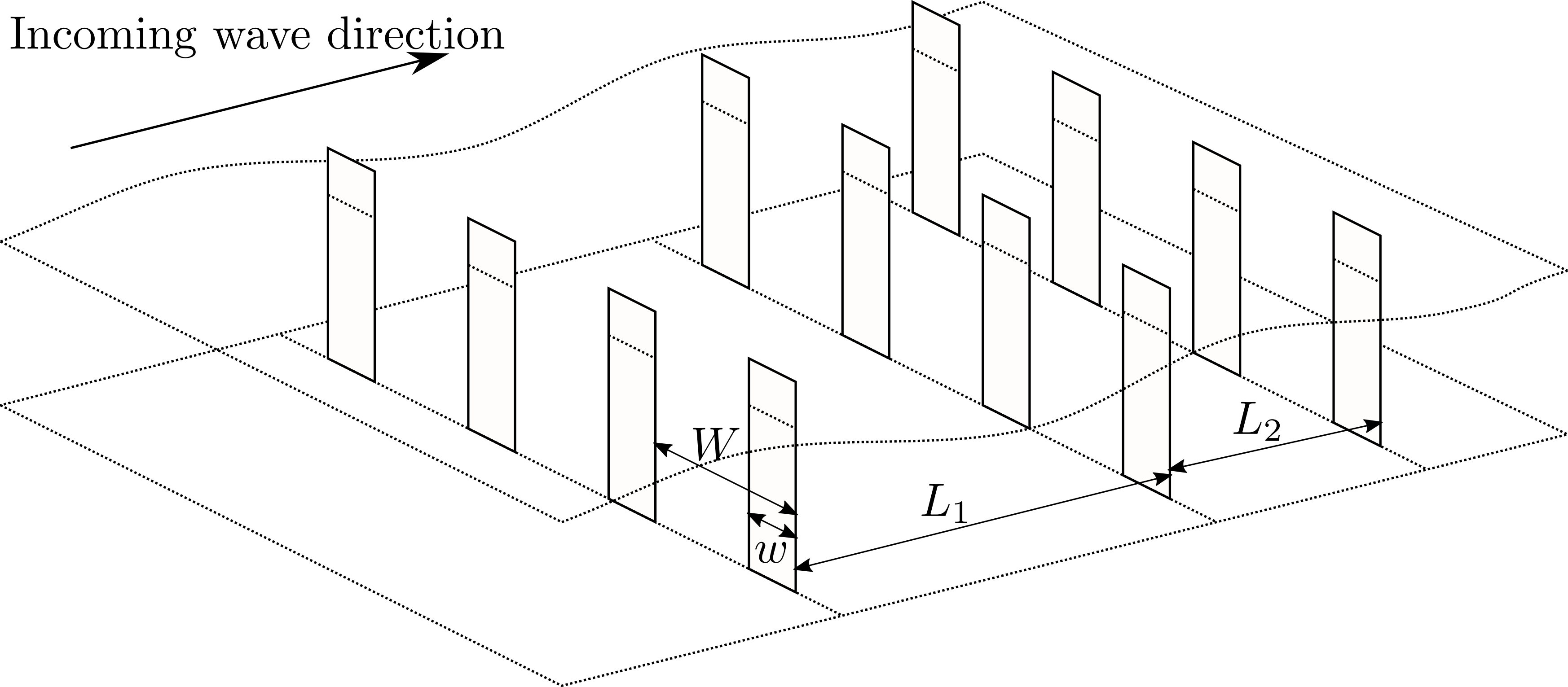}
    \caption{An array of wave absorbing vertical blades}
    \label{fig:blade_array}
\end{figure}

Consider an array of thin, vertical structures arranged in parallel rows, such as that illustrated in figure \ref{fig:blade_array}, where the extent of each row is infinite along the $y$ axis. The water depth is assumed constant and equal to 20 m. Linear potential flow theory is assumed.

The undisturbed incident wave is a plane wave propagating in the $x$ direction, orthogonal to the rows. The array is assumed periodic in the transverse $y$ direction, with periodicity smaller than the wavelength; in other words, $\lambda > W$. Therefore, there are no transverse modes in the flow velocity potential propagating away from a single row \cite{Dalrymple_1990}, and the waves propagating across consecutive rows can thus be represented as plane waves.

The incoming wave is described by means of the free-surface elevation (FSE), $\eta_0$, given by the following equation:\begin{equation} \label{eq:eta0}
    \eta_0(x,t) = \Re\{\hat{\eta}_0 e^{j(k x-\omega t)}\}
\end{equation} where $k=2\pi/\lambda$ is the wave number, $\omega$ is the wave frequency and $\hat{\eta}_0$ is the complex wave amplitude. As the incoming plane wave travels through a given row, the wave-row interaction results in a transmitted plane wave and a reflected plane wave, away from the row, due to the transverse periodicity smaller than $\lambda$. Because of the problem linearity, the transmitted and reflected complex wave amplitudes are linearly related to the incoming plane wave through (complex) transmission and reflection coefficients, respectively. 

Such a simple representation omits the evanescent modes in the row-to-row interaction process, which are essential to describe the flow in the close vicinity of each row of obstacles. The WSA implies, in particular, that the rows are sufficiently far from each other \cite{Evans_1990b}, although, as mentioned in the introduction, the latter assumption may not be strictly required in practice. Also note that the WSA does not assume that evanescent modes are nonexistent: it merely assumes that they can be neglected when analysing row-to-row interaction. The wave field, as computed through the wide-spacing approximation, is only a description of the flow outside the close vicinity of each row.

Finally, the interaction theory employed in this work does not assume identical rows: The obstacle characteristics, as well as their spacing, may vary \textit{across} rows, but not \textit{within a given row}; furthermore, the distance between consecutive rows may also vary - as illustrated in Fig. \ref{fig:blade_array}. However, for simplicity and results interpretation, the numerical case study in Sections \ref{s:casestudy} and \ref{s:results} will only consider identical, regularly-spaced rows.

\subsection{The case of fixed structures (diffraction problem)}
\label{ss:fixedstructure}

First consider the diffraction problem for a single row of (fixed) obstacles, located in $x=0$. The fixed row is subject to an incoming wave described as in \eqref{eq:eta0}. The reflected and transmitted waves can be written as follows: \begin{equation} \label{eq:rt_single} \begin{cases}
    \eta_{\text{t}}(x,t) = \Re\{\hat{t} \hat{\eta}_0 e^{j(kx - \omega t)}\}, \ \ x>0 \\
    \eta_{\text{r}}(x,t) = \Re\{\hat{r} \hat{\eta}_0 e^{j(-k x -\omega t)}\}, \ \ x< 0
    \end{cases}
\end{equation}

In \eqref{eq:rt_single}, $\hat{t}$ and $\hat{r}$ are the complex-valued transmission and reflection coefficients, which apply a change in amplitude and a phase shift to the incoming wave as it reaches the obstacle position. $\hat{t}$ and $\hat{r}$, for a fixed row, depend on the obstacle geometry, and on the incoming wave frequency. The row is symmetrical with respect to the $Oyz$ plane; therefore the reflection coefficient is identical for incoming waves propagating  in the positive and in the negative $x$ directions.

It is possible to state a few general properties which $\hat{r}$ and $\hat{t}$ must satisfy \cite{Linton_2011}, regardless of the precise row geometry. In linear wave theory, and considering no viscous dissipation at the interface with the obstacles, preservation of energy implies that the row reflection and transmission coefficients satisfy the following equality: \begin{equation} \label{eq:t2plusr2}
    |\hat{t}|^2+|\hat{r}|^2 = 1
\end{equation} 
Note that the energy-preservation property of \eqref{eq:t2plusr2}, should also be satisfied by the array as a whole. Considering thin rows with respect to the wave length, the following relation must also hold: \begin{equation}\label{eq:tplusr}
    \hat{t}+\hat{r}=1
\end{equation}
Indeed, one may think of the row, excited by two incoming waves of identical amplitude, and propagating in opposite directions. The two incoming waves together form a standing wave pattern. At the antinodes of the standing waves (i.e. at the locations where the wave amplitude is the largest), the horizontal fluid velocity is zero along the whole water column. Therefore, by synchronising the two exciting waves in such a way that the obstacle is at one such antinode, the no-flow boundary conditions on the obstacle vertical boundary are naturally satisfied. The presence of the obstacle thus leaves the flow unchanged. Formulated in terms of the reflection and transmission coefficients, this is simply written as $\hat{t}+\hat{r}= 1$. 

\begin{figure}
\begin{center}
    \includegraphics[width=0.5\linewidth]{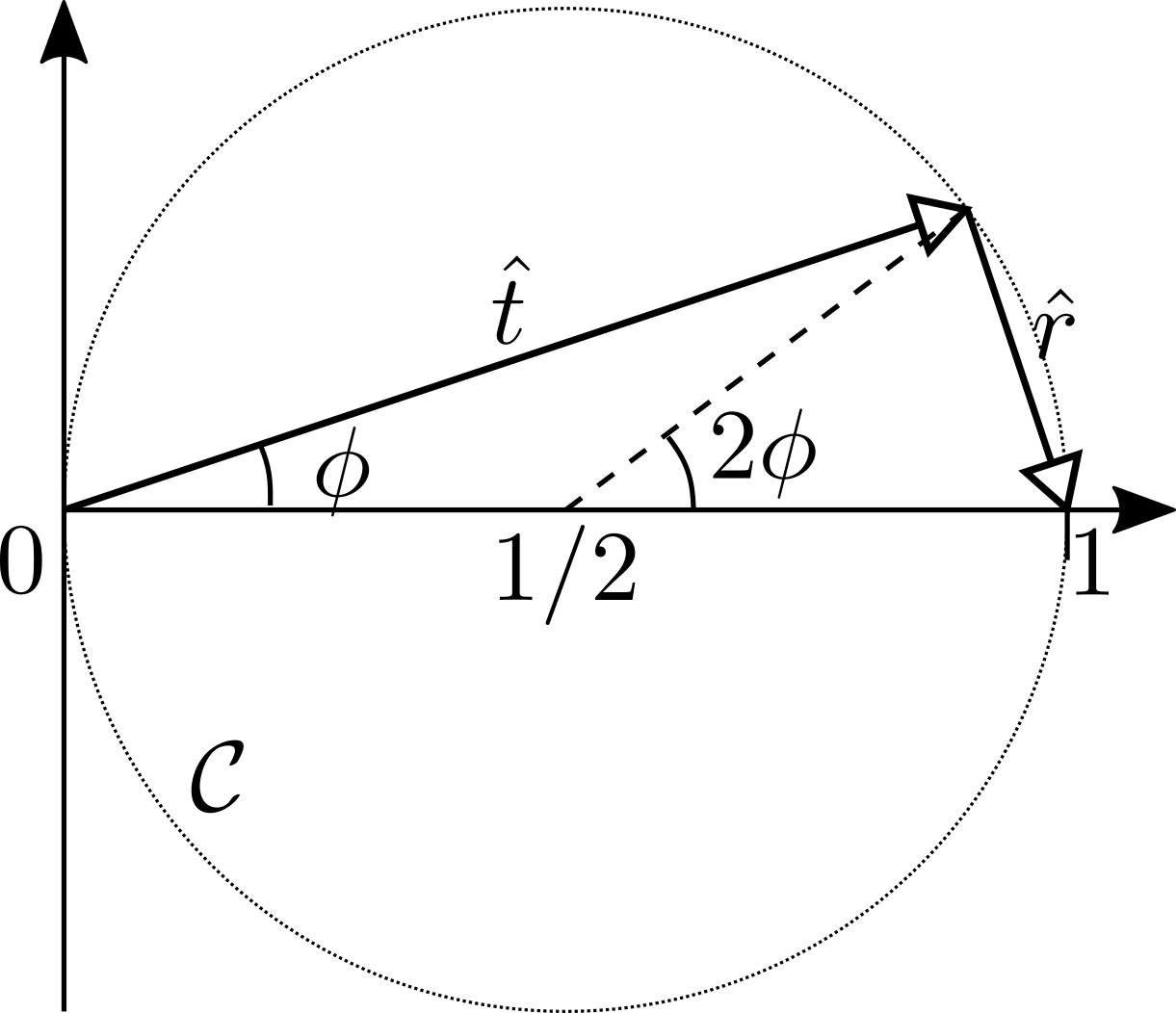}
  \caption{Geometrical characterisation of complex transmission and reflection coefficients for a row of obstacles held fixed.}
\label{fig:r_t}
\end{center}
\end{figure}

The properties represented by equations  \eqref{eq:t2plusr2} and \eqref{eq:tplusr} may be summarised geometrically, as shown in Fig. \ref{fig:r_t}. $\hat{t}$ and $\hat{r}$ can be visualised in the complex plane as forming two sides of a right triangle, of which the hypotenuse is of unitary length. In Fig. \ref{fig:r_t}, $\mathcal{C}$ denotes the circle, with centre $(1/2; 0)$ and radius $1/2$, on which $\hat{t}$ is located. Furthermore, relations \eqref{eq:t2plusr2} and \eqref{eq:tplusr} are equivalent to expressing $\hat{t}$ and $\hat{r}$ as a function of a single, real-valued parameter $\phi$, as follows: \begin{equation} \label{eq:ejphi}
\begin{cases}
    \hat{t} = \frac{1}{2}(1 + e^{2j\phi}) \\
    \hat{r} = \frac{1}{2}(1 - e^{2j\phi})
    \end{cases}
\end{equation} The parameter $\phi$, as illustrated in Fig. \ref{fig:r_t}, is in fact the angle of $\hat{t}$.

At this stage, no more assumptions are made on the specific structure geometry and lateral spacing (which would be required to know precisely where $\hat{t}$ is located on $\mathcal{C}$, as will be done in Section \ref{s:casestudy}). Furthermore, Fig. \ref{fig:r_t} corresponds to a single-frequency analysis. $\hat{t}$ and $\hat{r}$ are, in fact, frequency-dependent quantities; however, \eqref{eq:t2plusr2} and \eqref{eq:tplusr} ensure that $\hat{t}$ will remain on $\mathcal{C}$ for all frequencies, provided that the corresponding wavelength remains below the transverse row periodicity.

\subsection{The case of moving, wave-absorbing structures}
\label{ss:movingstructures}
\begin{figure}
\begin{center}
\includegraphics[width=0.7\linewidth]{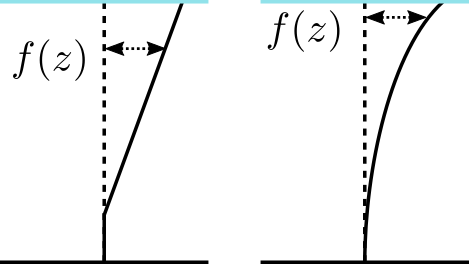}
  \caption{Examples of rigid mode (l.h.s.) and flexible mode (r.h.s.) deflection profiles}
\label{fig:owsc_deflection}
\end{center}
\end{figure}

Now consider that the vertical obstacles are allowed to move in a single mode of motion. More specifically, the mode of motion considered is any sort of deflection  in the $x$ direction, i.e. a deformation profile along the $z$ axis, of the form: \begin{equation} \label{eq:deflection}
    \xi(z,t) = f(z) \Re \{\hat{\xi} e^{-j\omega t} \}
\end{equation}
where $f(z)$ is the real-valued, unitary deflection profile of the vertical structures and $\hat{\xi}$ is the complex-valued motion amplitude. Formulation \eqref{eq:deflection} encompasses any rigid mode of motion in pitch or surge, as well as any mode of deformation in the case of flexible structures, as illustrated in Fig. \ref{fig:owsc_deflection}. Note that the present theory does not imply that obstacles are surface-piercing; generally they may occupy any fraction of the vertical extent between the sea bottom and the still water surface.

Let $\hat{A}_{+}$ and $\hat{A}_{-}$ be the two coefficients such that, for forced oscillations of the form \eqref{eq:deflection}, with complex \textit{velocity} amplitude $\hat{\nu} = -j\omega \hat{\xi}$, the wave radiated backward has amplitude $\frac{\omega}{g} \hat{A}_{-} \hat{\nu}$, and the wave radiated forward has amplitude $\frac{\omega}{g} \hat{A}_{+}\hat{\nu}$. Because of the motion anti-symmetry with respect to the $Oyz$ plane, $\hat{A}_{-}$ and $\hat{A}_{+}$ verify $\hat{A}_{-} = -\hat{A}_{+}$. 

The radiating characteristics $\hat{A}_{+}$ and $\hat{A}_{-}$ are related to the ``fixed-structure'' transmission and reflection coefficients $\hat{t}$ and $\hat{r}$, through Newman's relation \cite{Newman_1975}: \begin{equation} \label{eq:Newman}
    \hat{r}\hat{A}^*_{-} +    \hat{t}\hat{A}^*_{+} + \hat{A}_{-} = 0
\end{equation} where an asterisk $^*$ denotes complex-conjugate. Also using \eqref{eq:ejphi} and the fact that $\hat{A}_{-} = -\hat{A}_{+}$, equation \eqref{eq:Newman} eventually provides the following property regarding the phase of $\hat{A}_{+}$: \begin{equation}
    \hat{A}_{+} = -|\hat{A}_{+}|e^{j\phi}
\end{equation}

The application of Newton's second law to the structures oscillating in mode $f(z)$ yields the following frequency-domain dynamical equation: \begin{equation}
    \label{eq:dyn}
    (Z+Z_u)\hat{\nu} = \hat{h}_{\eta e} \hat{\eta_0}
\end{equation}
where: \begin{itemize}
    \item $\hat{h}_{\eta e}(\omega)$ denotes the complex coefficient which relates $\hat{\eta}$ to the excitation force.
    \item $Z$ denotes the complex impedance, defined as $Z(\omega) = B_{rad}(\omega) + j(K/\omega - \omega (I + A_{rad}(\omega)))$.
    \item $ B_{rad}(\omega) $ and  $A_{rad}(\omega)$ are, respectively, the frequency-dependent radiation damping and added mass coefficients.
    \item $K$ represents a stiffness coefficient, e.g. due to hydrostatic restoring forces (in the case of structures lighter than water), or the structure material flexural rigidity (in the case of a flexible structure). $I$ is the structure inertia for the mode of motion \eqref{eq:deflection}. Note that $K$ and $I$ depend on the internal properties of the chosen structures, and thus can be considered independent from the hydrodynamic problem.
    \item $Z_u = B_{u}(\omega) + j K_u(\omega)/\omega $ represents the effect of the PTO system, in the form of a possibly adjustable impedance, which consists of a combination of stiffness and damping terms.  
\end{itemize}  

First consider the structures moving freely under the effect of waves ($Z_u = 0$). Then the moving row transmission and reflection coefficients are calculated as the sum of the ``fixed-row'' (diffraction) and ``moving-row'' (radiation) effects : \begin{equation} \label{eq:TR} \begin{cases}
    \hat{T} = \hat{t} + \hat{A}_+ \frac{\omega}{g} \frac{\hat{h}_{\eta e}}{Z} = \frac{1}{2}(1 + e^{2j\phi} + 2 \hat{A}_+ \frac{\omega}{g}\frac{\hat{h}_{\eta e}}{B_{rad}} \frac{1}{1+j \gamma}) \\
    \hat{R} = \hat{r} - \hat{A}_+ \frac{\omega}{g} \frac{\hat{h}_{\eta e}}{Z} =  \frac{1}{2}(1 - e^{2j\phi} - 2 \hat{A}_+ \frac{\omega}{g}\frac{\hat{h}_{\eta e}}{B_{rad}} \frac{1}{1+j \gamma}) 
\end{cases}
\end{equation}
where $\gamma := \Im\{Z\} / B_{rad}$. It is clear that, in the absence of any dissipation or energy absorption mechanism at the structure level, and in the present potential flow framework, all the incoming wave energy is eventually transmitted or reflected; in other words, the mechanical energy provided by the waves to the structure motion is eventually given back to the flow in the form of radiated waves. Thus, $\hat{T}$ and $\hat{R}$ must satisfy a relation equivalent to \eqref{eq:t2plusr2}, that is, $|\hat{T}|^2+|\hat{R}|^2 = 1$. Together with \eqref{eq:TR}, the following condition ensues: \begin{equation} \label{eq:normcondition}
   \big| \big(1+j\gamma -  2 \hat{A}^*_+ \frac{\omega}{g}\frac{\hat{h}_{\eta e}}{B_{rad}} \big) \frac{1}{1+j \gamma}\big| = 1
\end{equation} Recalling that $\gamma$ may be chosen freely (since stiffness and inertia are independent from hydrodynamic considerations), \eqref{eq:normcondition} must hold for any $\gamma$, which implies the following condition: \begin{equation} \label{eq:GreenHaskind}
    \hat{A}^*_+ \frac{\omega}{g}\frac{\hat{h}_{\eta e}}{B_{rad}} = 1
\end{equation} Instead of the simple energetic considerations proposed above, \eqref{eq:GreenHaskind} can be retrieved by making use of the 2D Haskind relation, see \cite{Evans_1981}.

Finally, defining $\zeta = Z/B_{rad} = 1+j\gamma$, interesting expressions for $\hat{T}$ and $\hat{R}$ are obtained: \begin{equation} \label{eq:TRsimple} \begin{cases}
    \hat{T} = \frac{1}{2}(1 - \frac{\zeta^*}{\zeta} e^{2j\phi}) \\
    \hat{R} = \frac{1}{2}(1 + \frac{\zeta^*}{\zeta} e^{2j\phi}) 
\end{cases}
\end{equation} 

Defining $\zeta_u = Z_u/B_{rad}$, it is now easy to generalise \eqref{eq:TRsimple} to the case where some PTO force $Z_u \neq 0$ is applied: \begin{equation} \label{eq:TR_withU} \begin{cases}
    \hat{T} = \frac{1}{2}(1 - \frac{\zeta^*-\zeta_u}{\zeta+\zeta_u} e^{2j\phi}) \\
    \hat{R} = \frac{1}{2}(1 + \frac{\zeta^*-\zeta_u}{\zeta+\zeta_u} e^{2j\phi}) 
\end{cases}
\end{equation}

\begin{figure}
\begin{center}
\includegraphics[width=\linewidth]{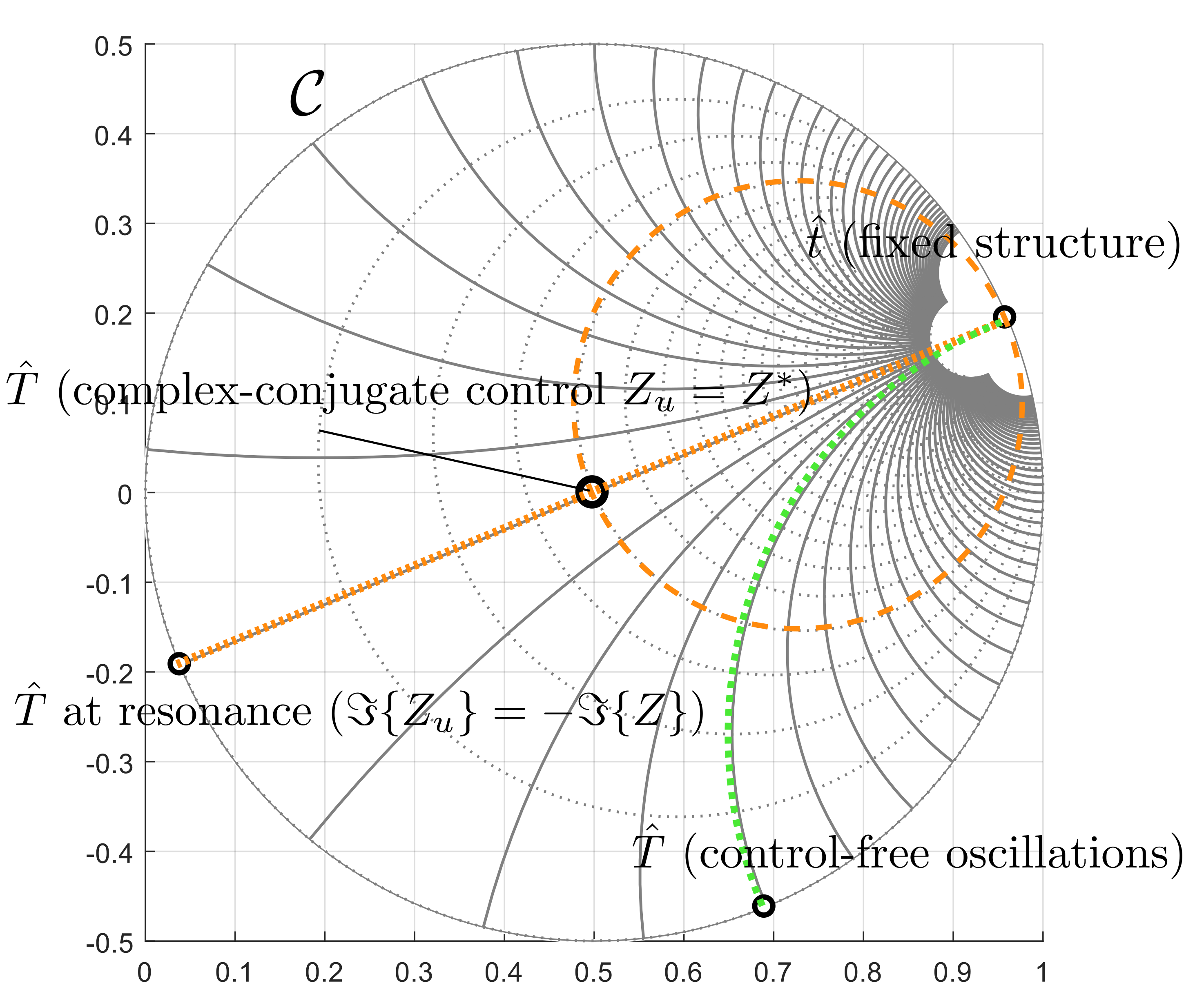}
  \caption{Position, in the complex plane, of the transmission coefficient of a row of vertical OWSCs, depending on its mechanical impedance. Each grey, dotted-line circle shows the locus of $\hat{T}$, for a fixed value of $B_u$, and $K_u = -\infty \rightarrow \infty$. Each grey, solid-line arc represents the locus of $\hat{T}$ for a fixed value of $K_u$, and $B_u = 0 \rightarrow \infty$. Through an appropriate choice of $B_u$ and $K_u$.}
\label{fig:r_t_moving}
\end{center}
\end{figure}

The expression for $\hat{T}$ in \eqref{eq:TR_withU} is illustrated in Fig. \ref{fig:r_t_moving}, where each grey, dotted-line circle shows the locus of $\hat{T}$ for a fixed value of $B_u$, and $K_u = -\infty \rightarrow \infty$, while each grey, solid-line arc represents the locus of $\hat{T}$ for a fixed value of $K_u$, and $B_u = 0 \rightarrow \infty$. By applying the appropriate resistive and reactive control, any $\hat{T}$ can be reached in the interior of $\mathcal{C}$. Additionally, defining $P_w$ and $P_a$ as the incoming wave and mechanically absorbed power values, respectively, it is easy to find the following expression for the power absorption ratio $\mu$: \begin{equation} \label{eq:mu}
    \mu = \frac{P_a}{P_w} = \frac{1-4d^2}{2}
\end{equation} where $d:=|\hat{T}-\frac{1}{2}|$ is the geometrical distance between $\hat{T}$ and the centre of $\mathcal{C}$.

Several special cases are interesting, some of which are illustrated by circular markers in Fig. \ref{fig:r_t_moving}: \begin{itemize}
    \item With no PTO damping ($B_u = 0$), all the energy remains in the flow; accordingly, $\hat{T}$ can only move along $\mathcal{C}$, and $\mu=0$. At resonance (either because the wave frequency is such that $\Im \{Z\} = 0$, or because the control impedance is such that $\Im \{Z_u\} = -\Im \{Z\}$), then $\hat{T} = \frac{1}{2}(1 - e^{2j\phi})$, i.e. $\hat{T}$ is the diametrical opposite of $\hat{t}$, the fixed-row transmission coefficient.
    \item A purely passive control ($K_u = 0$) can only drive $\hat{T}$ anywhere along the green arc.
    \item The well-known complex-conjugate control condition \cite{Falnes_2002} is $Z_u = Z^*$. In this case, one finds the well-known 2D energy absorption optimum for devices symmetric around the $Oyz$ plane, that is, $\mu = 1/2$. This corresponds to $\hat{T} = \hat{R} = 1/2$. 
    \item As either $K_u$ or $B_u$ grow to infinity, the structures become unable to move and therefore $\hat{T}\rightarrow \hat{t}$.    
\end{itemize}

In practice, for a specific geometry, solving for the diffraction problem indicates the position of $\hat{t}$ on $\mathcal{C}$. For freely-oscillating devices, the position of $\hat{T}_{Z_u = 0}$ on $\mathcal{C}$ ultimately depends on the relative weight of $B_{rad}$ and $\Im \{ Z\}$. Therefore, determining $\hat{T}_{Z_u = 0}$ is no longer a purely hydrodynamic problem, but depends both on the radiation problem solution (which is related to the deflection mode $f(z)$) \textit{and} on the specifics of the system considered, including inertia and internal stiffness, all encapsulated in $\Im \{Z\}$. 

In summary of this section, and leaving aside, for the moment, the practical difficulties that may arise with the application of an arbitrary, possibly non-causal PTO impedance $Z_u(\omega)$, \textit{there is, theoretically, a one-to-one relation between the impedance of the OWSC row (due to the combination of its hydrodynamic, mechanical and PTO/control characteristics) and its reflection-transmission-absorption properties. }

\section{Row-to-row interaction modelling using the wide-spacing approximation}

\label{s:wsa}

We follow, in this section, the model derived in \cite{Merigaud_2021}.

\subsection{General case}

Define a fluid domain $\mathcal{S} = [x;x']$ between two longitudinal positions $x$ and $x'$, comprising one or more rows of vertical structures such as those considered in Section \ref{s:singlerow}. Assume that the reflection and transmission problems have been solved for the domain $\mathcal{S}$, i.e. that complex coefficients $R_-$, $R_+$, $T_+=T_-=T$ have been found, such that, for incident wave components $\hat{\eta}_+$ and $\hat{\eta}'_-$ propagating into the domain $\mathcal{S}$, the wave components $\hat{\eta}_-$ and $\hat{\eta}'_+$, propagating away from $\mathcal{S}$, are derived as follows: \begin{equation} \label{eq:S} \begin{array}{ll}
    \hat{\eta}_- &  = R_+ \hat{\eta}_+ + T \hat{\eta}'_- \\
    \hat{\eta}'_+ & = R_- \hat{\eta}'_- + T \hat{\eta}_+
\end{array}
\end{equation} In the above expression, the ``forward'' and ``backward'' transmission coefficients, $T_+$ and $T_-$, are assumed identical, which will receive proper justification subsequently. 

\begin{figure}
    \centering
    \includegraphics[width = 0.6\columnwidth]{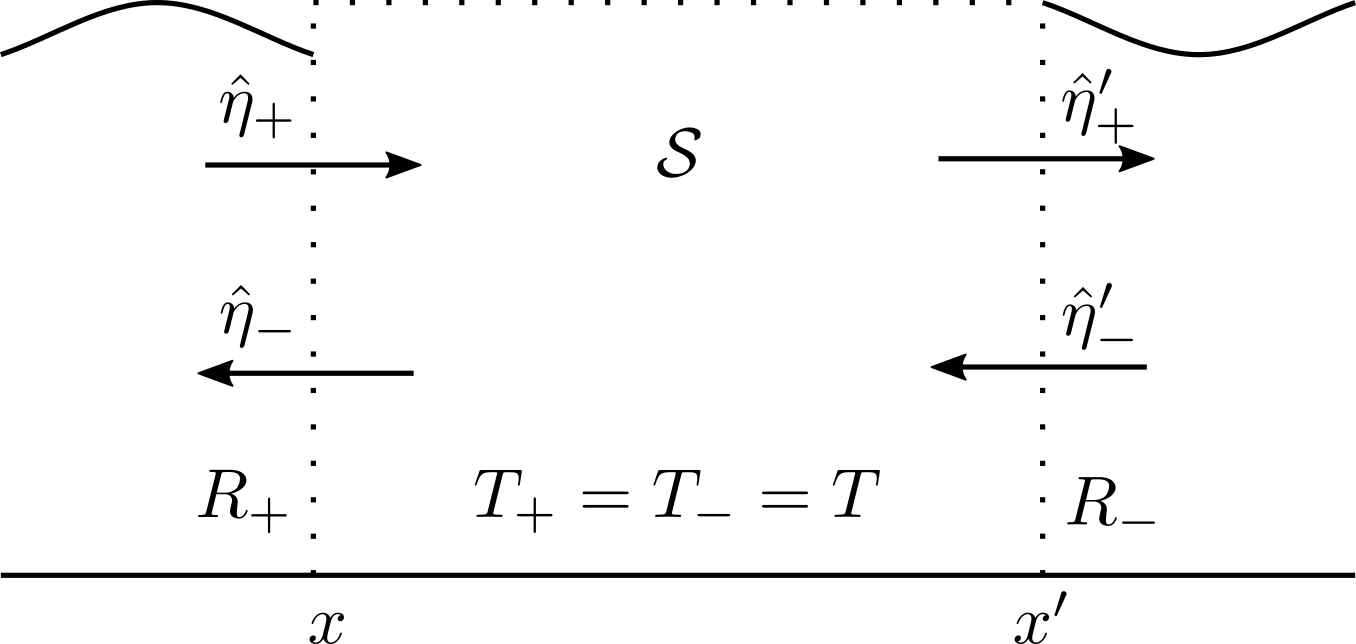}\\ 
    \includegraphics[width = 0.6\columnwidth]{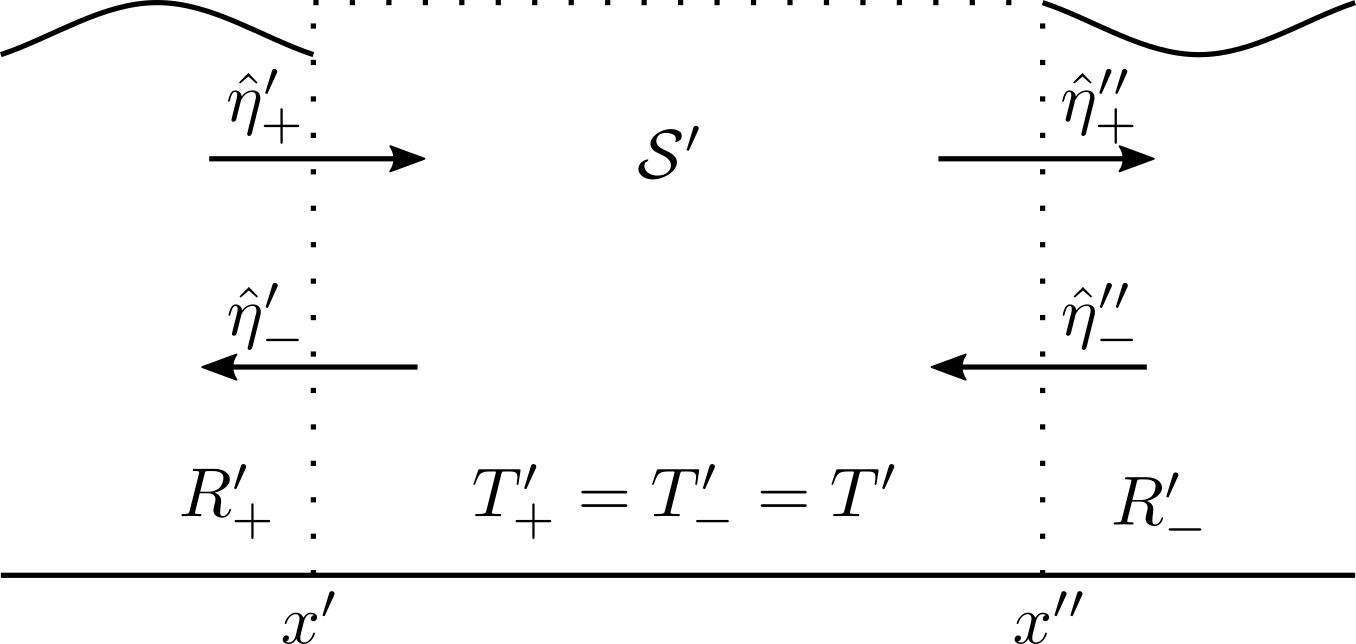}\\ \vspace{3mm}
    \includegraphics[width = 0.9\columnwidth]{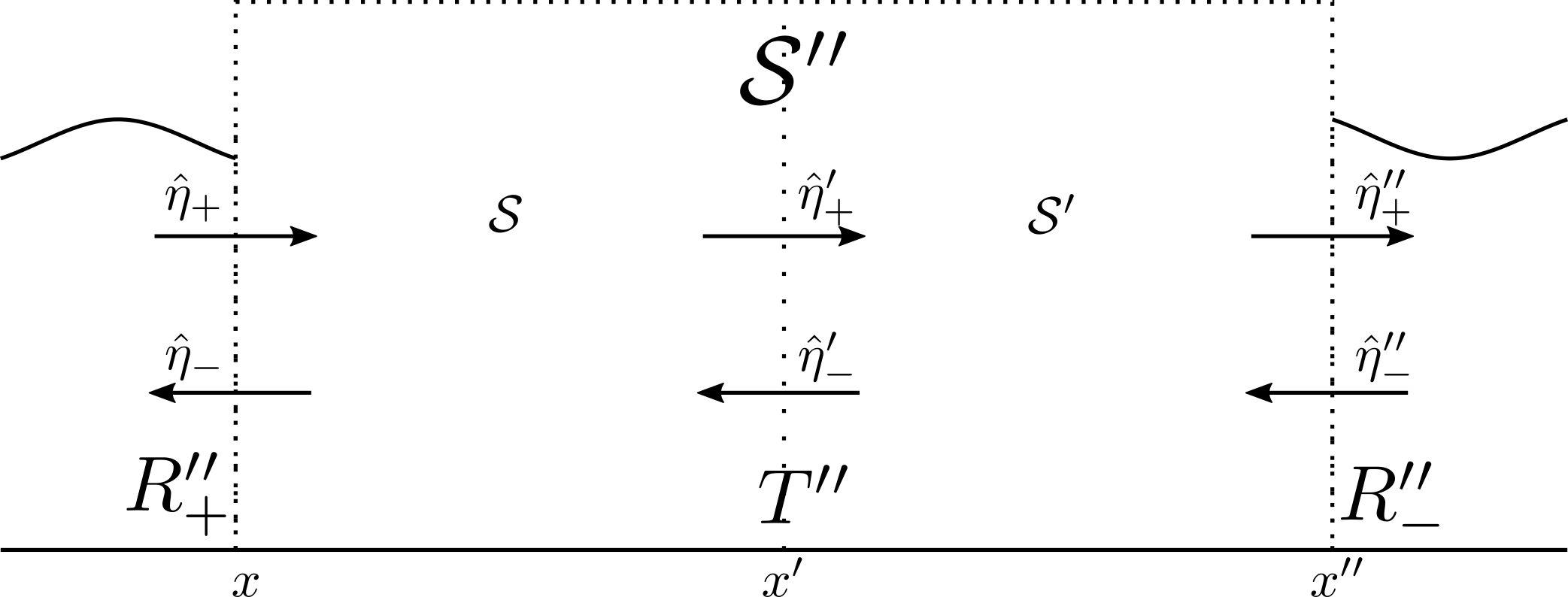}
    \caption{Wave transmission and reflection through a domain $\mathcal{S}$ (top), a domain $\mathcal{S}'$ (middle), and a domain $\mathcal{S}''$ that is the combination of $\mathcal{S}$ and $\mathcal{S}'$ (bottom) (figure reproduced from \cite{Merigaud_2021}).}
    \label{fig:S}
\end{figure}

Similarly, let $\mathcal{S}'$ be the domain comprised between $x'$ and another position $x'' > x'$, and also represented in Fig. \ref{fig:S}. Complex coefficients $R'_-$, $R'_+$, $T'$ have been found, such that, for incident wave components $\hat{\eta}'_+$ and $\hat{\eta}''_-$, the wave components $\hat{\eta}'_-$ and $\hat{\eta}''_+$ are determined as follows: \begin{equation} \label{eq:Sprime} \left. \begin{array}{ll}
    \hat{\eta}'_- & = R'_+ \hat{\eta}'_+ + T' \hat{\eta}''_- \\
    \hat{\eta}''_+ & = R'_- \hat{\eta}''_- + T' \hat{\eta}'_+
\end{array}\right\}
\end{equation}

Now define $\mathcal{S}''$ as the domain extending from $x$ to $x''$, as represented in figure \ref{fig:S}. By combining \eqref{eq:S} and \eqref{eq:Sprime}, it is straightforward to obtain a linear relation between $\{\hat{\eta}_+, \hat{\eta}''_-\}$, on the one hand, and $\{\hat{\eta}_-, \hat{\eta}''_+\}$ on the other hand, such that: \begin{equation} \label{eq:Sprimeprime}  \begin{cases}
    \hat{\eta}_- = R''_+ \hat{\eta}_+ + T'' \hat{\eta}''_- \\
    \hat{\eta}''_+ = R''_- \hat{\eta}''_- + T'' \hat{\eta}_+
\end{cases}
\end{equation} where the coefficients $R''_+$, $R''_-$ and $T''$ are calculated as follows: \begin{equation}\label{eq:recursion_general}  \begin{cases}
    R''_{+}=\frac{R_{-} - R'_{-}(R_{+}R_{-}-T^2)}{1-R'_{-}R_{+}} \\
    R''_{-}=\frac{R^{'}_{+} - R_{+}(R'_{+}R'_{-}-T'^{2})}{1-R'_{-}R_{+}}\\
    T''_+ = T''_- =T'' = \frac{T'T}{1-R'_{-}R_{+}}
    \end{cases}
\end{equation} 
Note that, in the above expression, if both $\mathcal{S}$ and $\mathcal{S}'$ satisfy the condition that their forward and backward transmission coefficients are identical, the same holds for $\mathcal{S}''$.   

For an array comprising $N$ rows in positions $X_1... X_N$, recursion \eqref{eq:recursion_general} can be put into practice by defining the elementary domains $\mathcal{S}_n$ for $n = 1...N$, only containing the $n^{th}$ row, and extending from $x_{n-1}$ to $x_{n}$, where, for $n = 1...N-1$, $x_{n}$ is halfway between $X_n$ and $X_{n+1}$, and $x_0$ and $x_N$ are arbitrary longitudinal positions upwave and downwave relative to the array, respectively. For one such domain, it is easy to find that the transmission and reflection coefficients are as follows: \begin{eqnarray} 
    T_n = \hat{t}_n e^{i k(x_{n}-x_{n-1})} \\
    R_{n+} = \hat{r}_n e^{2 i k(X_{n}-x_{n-1})} \\
    R_{n-} = \hat{r}_n e^{2 i k(x_{n}-X_{n})} 
\end{eqnarray} The recursion can be initiated using the elementary domain containing the first row, thus obtaining $T_1$, $R_1^+$ and $R_1^-$. Additional rows are then added sequentially through the use of \eqref{eq:recursion_general}. It can be seen that, for every elementary domain, the forward and backward transmission coefficients are identical, and that this property is preserved through successive iterations of \eqref{eq:recursion_general}, which justifies \textit{a posteriori} the simplifications $T_+ = T_-$ in \eqref{eq:S} and \eqref{eq:Sprime}.

\subsection{The special case of identical rows}

In the special case where all rows of obstacles are identical (i.e. $\hat{T}$ is identical for every row) and regularly spaced with spacing $L$, analytical expressions for the reflection and transmission coefficients of an $N$-row array can be found, see \cite{Evans_1990b} and \cite{Merigaud_2021}. From those expressions, it is possible to derive asymptotic formulae for the case where the number of rows grows to infinity. Only those asymptotic formulae are reproduced in the following, because they will be employed as a limiting case in the numerical experiments considered in Sections \ref{s:casestudy} and \ref{s:results}.

Using the results of Section \ref{s:singlerow}, parametrise $\hat{T}$ by means of two real numbers $\psi$ and $\rho$ such that:  \begin{equation}
    \hat{T} = \frac{1}{2}+\rho e^{2j\psi}, \ 0\leq \rho \leq \frac{1}{2}
\end{equation} For any angle $\theta$, define the complex function $g$ as follows: \begin{equation}
    g(\theta) = \frac{1}{2}e^{-i\theta} - \rho e^{i \theta}
\end{equation}
Finally, defining $\beta$ such that $\cosh(\beta)=\frac{g(k L + \psi)}{g(\psi)}$, the infinite array reflection coefficient can be expressed as follows: \begin{equation}
R_{\infty} =e^{-\beta}    
\end{equation} 
Note that $R_{\infty}$ and all expressions above are frequency-dependent, because $\hat{T}$ (and thus $\psi$ and $\rho$) as well as the wavenumber $k$, are all frequency-dependent.

\section{Numerical case study}
\label{s:casestudy}

\begin{figure}
\begin{center}
\includegraphics[width=0.9\linewidth]{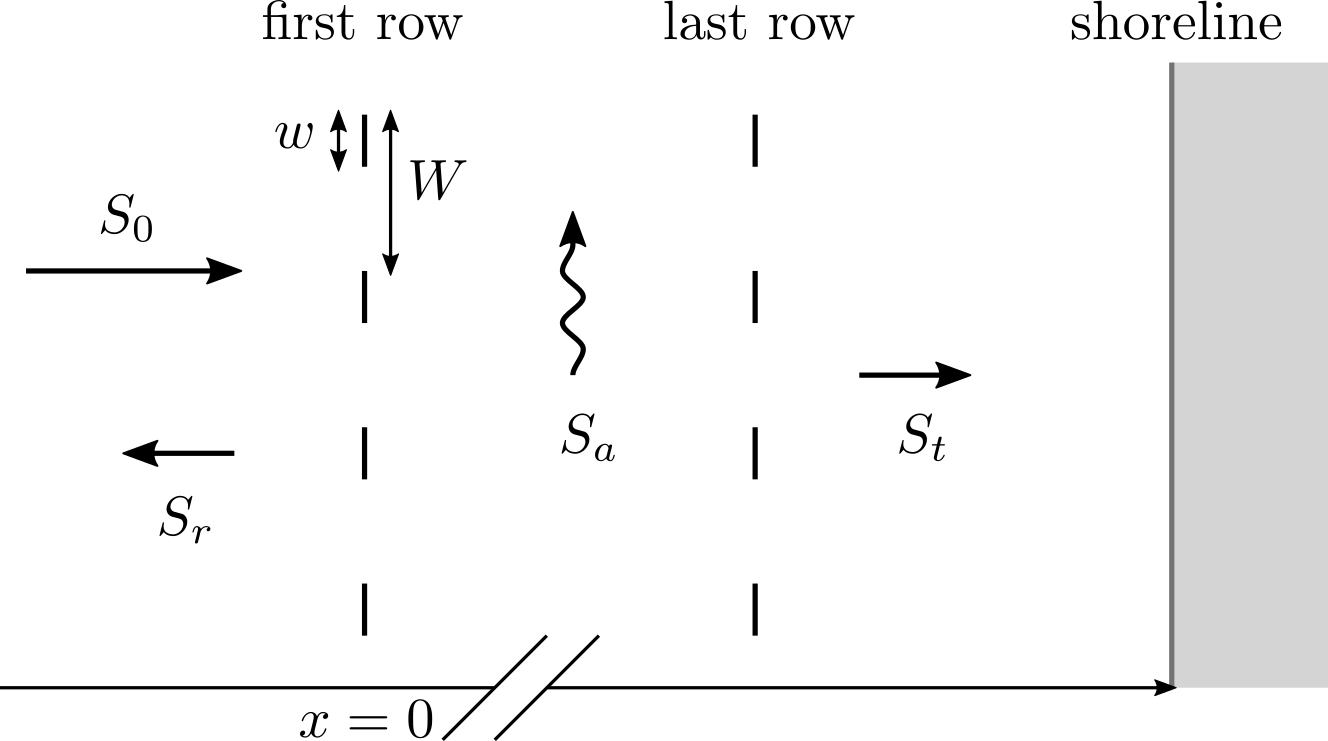}
  \caption{Top view of the numerical case study layout and main notations}
\label{fig:numsetup}
\end{center}
\end{figure}

In this example numerical case study, arrays of bottom-mounted, surface-piercing, vertical square plates, such as those illustrated in Fig. \ref{fig:blade_array}, are considered. The geometry of a given row of such structures is therefore characterised by two real numbers $w$ and $W$, where $w$ is the structure width and $W$ is the row transverse periodicity, as specified in Fig. \ref{fig:blade_array}. Figure \ref{fig:numsetup} shows a top view of the problem considered, where, without loss of generality, the origin $x=0$ is the first row longitudinal position, and there can be any number of rows between the first and last rows. 

The diffraction problem for a single row is the well-known ``slotted barrier'' problem in linear hydrodynamics. Analytical formulae for the reflection and transmission coefficients are determined, based on \eqref{eq:tplusr} and the following result, given in \cite{Dalrymple_1990}: \begin{equation} \label{eq:dalrymple}
    \frac{\hat{r}(\omega)}{1-\hat{r}(\omega)} = \sum\limits_{m=1}^{\infty} \frac{2k(\omega) J_0^2(m \pi  \frac{W-w}{W})}{\sqrt{k^2(\omega)-(\frac{2m\pi }{W})^2}}
\end{equation} where $J_0$ denotes a Bessel function, and the sum is truncated to $M = 10^5$. Equation \eqref{eq:dalrymple} depends on two ratios: transverse periodicity to wavelength, and  aperture to transverse periodicity. Reflection is a monotonic, increasing function of $W/\lambda$. It can be verified that the coefficients $\hat{t}$ and $\hat{r}$, thus obtained, do satisfy the geometrical relations highlighted in Section \ref{s:singlerow}.

\begin{figure}
\begin{center}
\includegraphics[width=\linewidth]{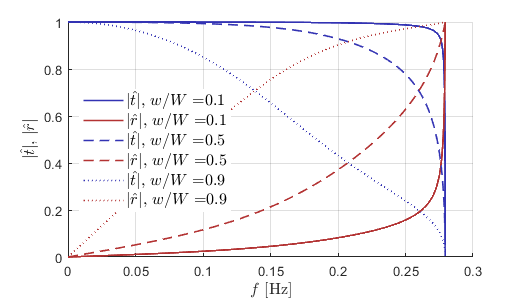}
  \caption{Complex transmission and reflection coefficient magnitude of a row of vertical, fixed structures}
\label{fig:rt_fixed}
\end{center}
\end{figure}

In this particular study, the transverse periodicity is chosen to be $W=20$ m. Thus, the simple reflection/transmission results of Section \ref{s:wsa} remain valid for wave frequencies below $f_c=0.28$ Hz (such that the wavelength equals $W$), above which transverse propagating components would complicate the analysis. $f_c$ is sufficient to cover the vast majority of the energy contained in typical swell spectra, with peak wave periods in the range of $8-20$s \cite{Babarit_2017_Book}. The norms of the fixed-structure transmission and reflection coefficients $\hat{r}$ and $\hat{t}$ are shown in Fig. \ref{fig:rt_fixed}, over the range of frequencies considered. As can be appreciated in Fig. \ref{fig:rt_fixed}, individual rows are predominantly transmissive (except for frequencies approaching $f_c$), unless the blockage ratio $w/W$ is close to unity, which would represent a large barrier with tiny apertures. The latter case, however, would not be consistent with the underlying idea of this work, which rather consists of exploiting array interaction effects to absorb and reflect wave energy, without resorting to large structures.

The incident flow is a Gaussian, long-crested wave field, characterised by a JONSWAP \cite{Hasselmann_1973} spectral density function $S_0$, with peak period $T_p = 12$ s, significant height $H_s =1$ m and peak enhancement factor $\gamma = 2$.

The case of fixed rows is first considered, described by $\hat{r}$ and $\hat{t}$ determined from \eqref{eq:dalrymple} as explained above. Then, oscillating rows (i.e. OWSCs) are investigated. In order to avoid going into the specifics of the devices mode of operation, the generic considerations detailed in Section \ref{ss:movingstructures} are utilised, by considering two control strategies: \begin{itemize}
    \item In the ``complex-conjugate control'' (CCC) strategy, every row of OWSCs is assumed to operate under ideal complex-conjugate control, which corresponds to $\hat{T} = \hat{R} = 1/2$. The hydrodynamic efficiency is $\mu = 1/2$ for a single row of devices.
    \item The ``over-damped control'' (ODC) strategy is a modification of the CCC strategy, whereby the applied PTO damping is 3 times larger ($B_u = 3 B_{rad}$), which is easily shown to correspond to $\hat{T}(\omega) = \frac{1}{2}(\hat{t}(\omega)+1/2)$ (visually in Fig. \ref{fig:r_t_moving}, $\hat{T}$ would be halfway between $1/2$ and $\hat{t}$). With ODC, using \eqref{eq:mu}, the hydrodynamic efficiency is $\mu = 3/8$ for a single row of devices.
\end{itemize}

Assuming $N$ identical rows and homogeneous row-to-row spacing, the recursion of Section \ref{s:wsa} is employed frequency-wise to obtain the array reflection and transmission coefficients, $R_N(\omega)$ and $T_N(\omega)$, using single-row coefficients as building blocks: $\{\hat{r},\hat{t}\}$ in the case of fixed rows, $\{\hat{R},\hat{T}\}$ for (moving) rows operating under the two control strategies outlined above. The asymptotic case $N\rightarrow \infty$ is also considered. The shoreline is assumed to absorb all the energy that reaches it, i.e. it does not reflect any wave back to the array. However, the effect of total or partial wave reflection by the shore would be easily included in the analysis, using the recursion of Section \ref{s:wsa}.

Of particular interest are the reflected ($S_r$), transmitted ($S_t$) and absorbed ($S_a$) wave spectra (the latter being simply calculated as the part of the wave spectrum which is neither transmitted, nor reflected, since in this study losses are not considered). Furthermore, in the up-wave zone ($x < 0$), reflected and incident waves together form interaction patterns, which induce spatial inhomogeneity of the total wave spectrum $S_{tot}(x,\omega)$. All aforementioned spectra are calculated from $S_0$ and $\{R_N,T_N\}$ as follows: \begin{equation} \label{eq:spectra} \small \begin{cases}
    S_r(\omega) = |R_N(\omega)|^2 S_0(\omega) \\
    S_t(\omega) = |T_N(\omega)|^2 S_0(\omega) \\
    S_a(\omega) = (1-|R_N(\omega)|^2-|T_N(\omega)|^2) S_0(\omega) \\
   \begin{split}  S_{tot}(x,\omega) &=  \big[1 + |R_N(\omega)|^2 \\
   &+ 2\Re \{R_N(\omega) e^{-2j k(\omega)x}\}\big]S_0(\omega), \ \ x<0 \end{split}
\end{cases}
\end{equation}

\section{Results}
\label{s:results}
\subsection{Fixed structures}

\label{ss:res_fixed}

\begin{figure}[h]
  \begin{center}
  \subfloat[\scriptsize $N=5$, $\frac{w}{W} = 0.1$]{
      \includegraphics[width=0.33\columnwidth]{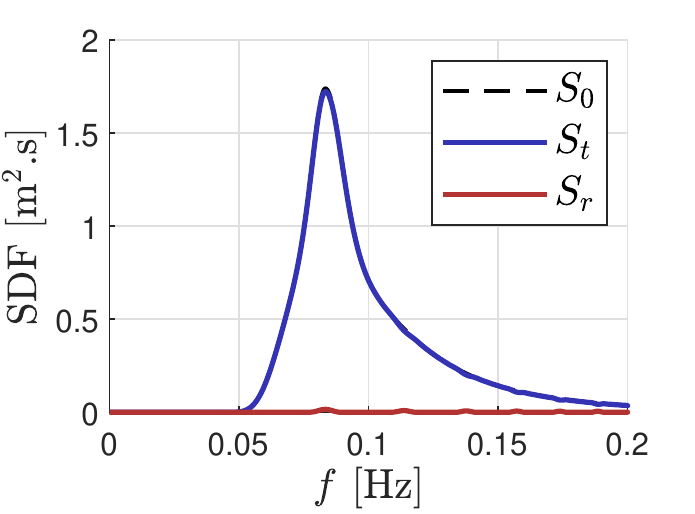}
      \label{sub:N5b10} 
      } 
    \subfloat[\scriptsize  $N=5$, $\frac{w}{W} = 0.5$]{
      \includegraphics[width=0.33\columnwidth]{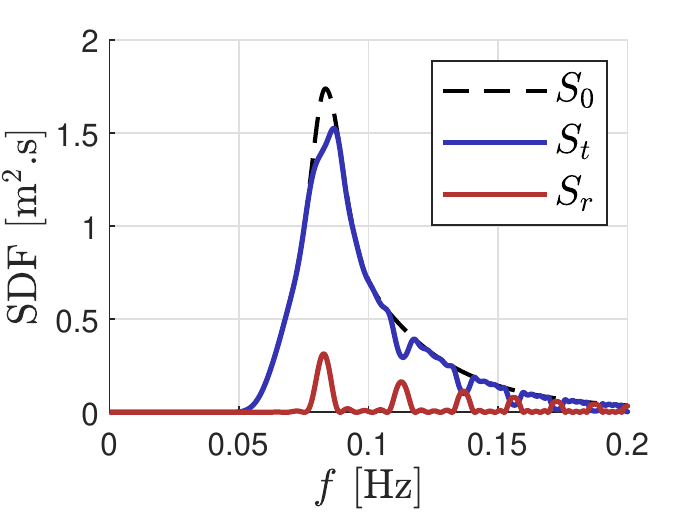}
      \label{sub:N5b50}
                         } 
    \subfloat[\scriptsize $N=5$, $\frac{w}{W} = 0.8$]{
      \includegraphics[width=0.33\columnwidth]{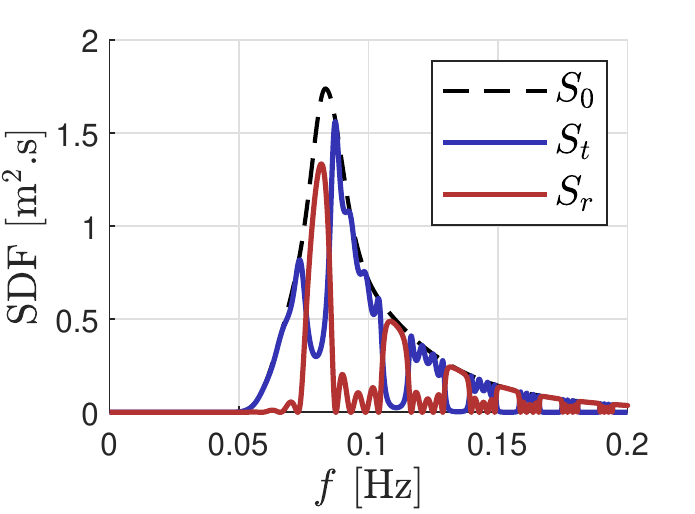}
      \label{sub:N5b80}
                         } \\
     \subfloat[\scriptsize $N=10$, $\frac{w}{W} = 0.1$]{
      \includegraphics[width=0.33\columnwidth]{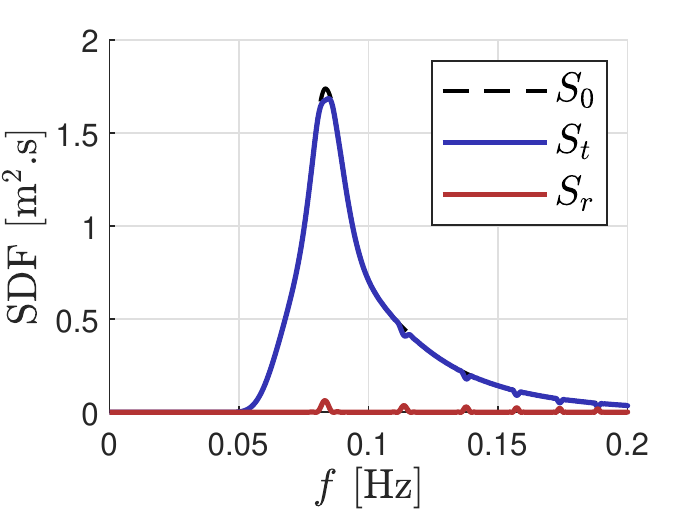}
      \label{sub:N10b10} 
      } 
     \subfloat[\scriptsize $N=10$, $\frac{w}{W} = 0.5$]{
      \includegraphics[width=0.33\columnwidth]{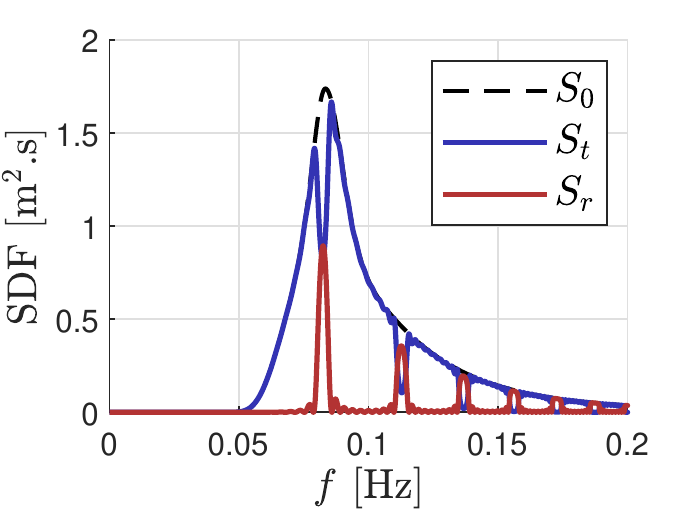}
      \label{sub:N10b50}
                         } 
    \subfloat[\scriptsize $N=10$, $\frac{w}{W} = 0.8$]{
      \includegraphics[width=0.33\columnwidth]{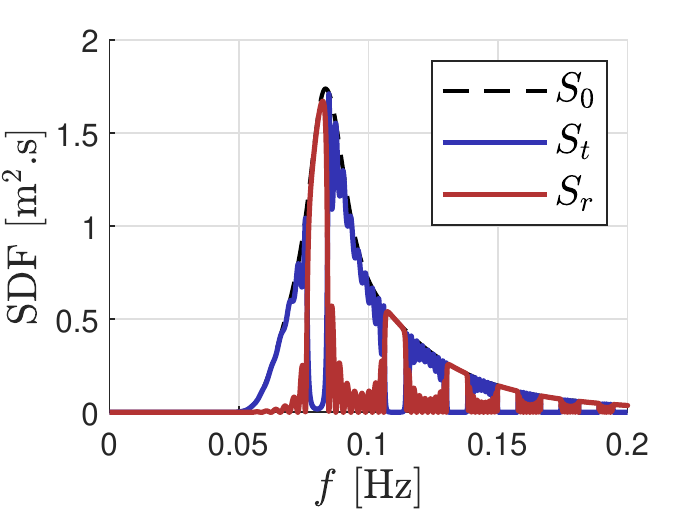}
      \label{sub:N10b80}
                         } \\
     \subfloat[\scriptsize $N\rightarrow \infty$, $\frac{w}{W} = 0.1$]{
      \includegraphics[width=0.33\columnwidth]{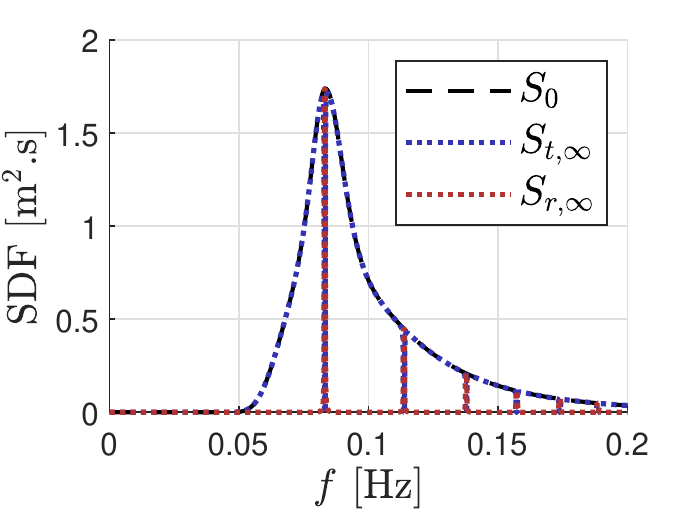}
      \label{sub:NinfB10} 
      }  
     \subfloat[\scriptsize $N\rightarrow \infty$, $\frac{w}{W}= 0.5$]{
      \includegraphics[width=0.33\columnwidth]{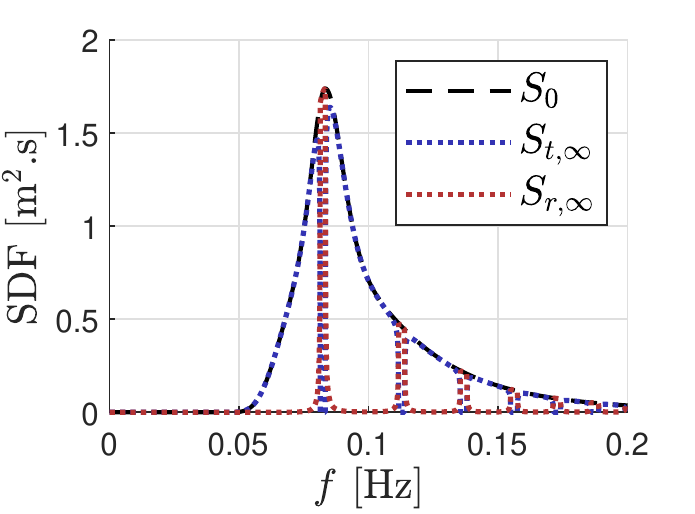}
      \label{sub:NinfB50}
                         } 
    \subfloat[\scriptsize $N\rightarrow \infty$, $\frac{w}{W}= 0.8$]{
      \includegraphics[width=0.33\columnwidth]{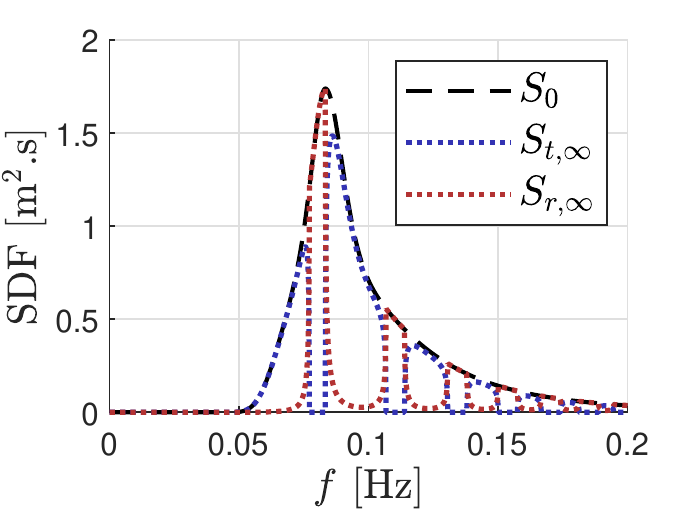}
      \label{sub:NinfB80}
                         }
    \caption{Incident, transmitted and reflected spectra for fixed structures, in various array configurations and blockage ratios, keeping $L = \lambda_p$}
    \label{fig:results_fixed}
  \end{center}
\end{figure}

In an array of fixed structures, no energy is absorbed, so that the incident energy is solely split into transmitted and reflected energy. With a view to coastal protection, the objective here is to determine whether array interaction effects can effectively mitigate wave transmission. The array transmission and reflection coefficients are computed for blockage ratios $w/W = 0.5$ and $0.8$, and different numbers of rows $N = 5, 10$ and $N\rightarrow \infty$. The row-to-row spacing $L$ is set to $\lambda_p$, the wavelength at the peak wave period. The resulting reflected and transmitted spectra are plotted in Fig. \ref{fig:results_fixed}, together with the incident spectrum $S_0$.

Interaction effects are clearly visible in Fig. \ref{fig:results_fixed}, in the form of the well-known \textit{band-gap} intervals, where reflection is higher, see e.g. \cite{Linton_2011}. Those occur in the vicinity of Bragg frequencies $f^*_n$ such that $2L = n\lambda(f^*_n), \ n = 1, 2, ...$, where $\lambda(f^*_n)$ denotes the wavelength which satisfies the dispersion relation with $f^*_n$. With a large number of rows ($N\rightarrow \infty$), reflection is total within each band-gap interval. With a finite number of rows, reflection exhibits oscillatory behaviour between consecutive band-gaps intervals. 

Furthermore, it can be shown \cite{Linton_2011,Merigaud_2021} that the band-gap interval width is governed by the magnitude of the parameter $\phi \in [0; \pi/2]$ in \eqref{eq:ejphi}, which represents the balance between the reflection and transmission properties of individual rows. The latter result is exemplified in Fig. \ref{fig:results_fixed}, through the differences between the results of the two blockage ratios, where the larger value of $w/W$ yields wider band-gap intervals. Yet, it is striking to observe that, even with the higher ratio $w/W = 0.8$, which corresponds to 16m-wide structures separated by 4m-wide apertures, the vast majority of the incoming energy is transmitted to the shore, while interaction effects can merely mitigate transmission within frequency intervals, too narrow to effectively protect the coastline.

\subsection{Oscillating structures}
\label{ss:res_mov}
\begin{figure}[h]
  \begin{center}
    \subfloat[$N=1$, $\Re \{Z_u\}=B_{rad}$]{
      \includegraphics[width=0.5\columnwidth]{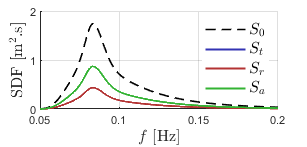}
      \label{sub:movN1od0}
                         }
    \subfloat[$N=1$, $\Re \{Z_u\} =3 B_{rad}$]{
      \includegraphics[width=0.5\columnwidth]{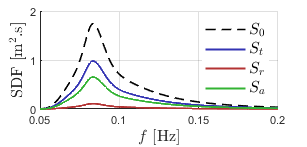}
      \label{sub:movN1od50}
                         } \\
     \subfloat[$N=5$, $\Re \{Z_u\} = B_{rad}$]{
      \includegraphics[width=0.5\columnwidth]{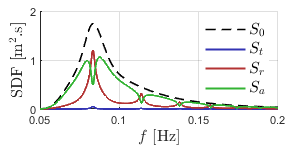}
      \label{sub:movN5od0}
                         }
    \subfloat[$N=5$, $\Re \{Z_u\} =3 B_{rad}$]{
      \includegraphics[width=0.5\columnwidth]{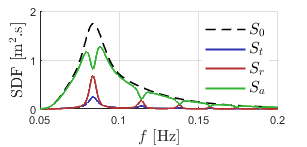}
      \label{sub:movN5od50}
                         } \\
     \subfloat[$N\rightarrow \infty$, $\Re \{Z_u\} = B_{rad}$]{
      \includegraphics[width=0.5\columnwidth]{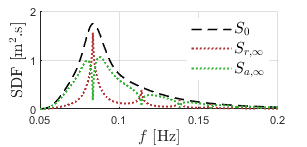}
      \label{sub:movNinfod0}
                         }
    \subfloat[$N\rightarrow \infty$, $\Re \{Z_u\} =3 B_{rad}$]{
      \includegraphics[width=0.5\columnwidth]{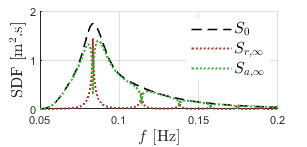}
      \label{sub:movNinfod50}
                         }
    \caption{Incident, transmitted and reflected spectra for OWSCs operating under the CCC (l.h.s.) and ODC (r.h.s.) strategies, in various array layout configurations, keeping $L = \lambda_p$}
    \label{fig:results_mov}
  \end{center}
\end{figure}

With oscillating structures, i.e. OWSCs, the incident energy is split into transmitted, reflected and absorbed energy. In order to explore how relatively small structures may efficiently reflect or absorb waves, the width of each device is set to $w=2$ m only ($w/W = 0.1$). The corresponding spectra are shown in Fig. \ref{fig:results_mov}, for various numbers of rows, and for the two control strategies introduced in Section \ref{s:casestudy}.

First consider the CCC strategy (l.h.s. graphs of Fig. \ref{fig:results_mov}). The case $N = 1$ corresponds to the well-known 2D optimal power absorption by a single WEC symmetric around the $Oyz$ plane: the absorbed power is half of the incoming power ($\mu = 1/2$), while the remaining energy is evenly shared into transmission and reflection (which is why $S_t$ is hidden by $S_r$ in Fig. \ref{sub:movN1od0}). As more rows are added, each additional row absorbs a fraction of the energy it receives, so that for large numbers of rows, the array transmission tends to zero. The asymptotic case $N\rightarrow \infty$ is a satisfactory approximation to smaller array behaviour, even for as little as $N=5$ rows, see Figs. \ref{sub:movN5od0} and \ref{sub:movNinfod0}.

Most interestingly, array interaction patterns are clearly visible in Fig. \ref{fig:results_mov}, but they now take the form of marked reflection peaks (as opposed to band-gap intervals), which occur precisely at the Bragg resonant frequencies $f^*_n$, while, in the frequency intervals between successive Bragg peaks, incoming energy is predominantly absorbed. Note that, in the present case, since $L = \lambda_p$, the peak wave period exactly corresponds to the second Bragg peak location.

ODC results are now examined (r.h.s. of Fig. \ref{fig:results_mov}). As can be appreciated by comparing Fig. \ref{sub:movN1od50} with Fig. \ref{sub:movN1od0}, ODC is sub-optimal from a single-row power absorption perspective. More specifically, from \eqref{eq:mu} the corresponding absorption efficiency is $\mu = 3/8$ instead of $\mu = 1/2$. In contrast, for arrays comprising several or many rows, ODC becomes preferable to CCC for energy harvesting. Intuitively formulated, in ODC, the magnitude of $\hat{T}$ increases, while that of $\hat{R}$ decreases, relatively to CCC. Therefore, front rows reflect less energy back to the ocean (which would be lost for harvesting purposes) while transmitting more energy for subsequent rows to harvest. However, too large a PTO damping value would ultimately prevent any motion from occurring (see the location of $\hat{t}$  in Fig. \ref{fig:r_t_moving}), thus yielding the same results as in the ``fixed-row'' case of Section \ref{ss:res_fixed}. Those considerations suggest that non-trivial optimal control tuning parameters are yet to be determined, informed by the array size and layout. 

Finally, it is interesting to assess how the OWSC array affects the wave field in its surroundings. Concerning the area between the array and the shoreline, Fig. \ref{fig:results_mov} indicates that, for a sufficiently large number of rows, both control strategies offer close-to-perfect coastal protection, with almost zero wave energy transmitted.

\begin{figure}
\begin{center}
\includegraphics[width=\linewidth]{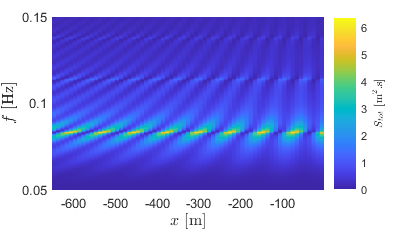}
  \caption{$S_{tot}(x,f)$ in the up-wave zone, where $x$ denotes the longitudinal position relative to the first row. Infinite array under the CCC strategy, $L = \lambda_p$}
\label{fig:Stot}
\end{center}
\end{figure}

In contrast, in both strategies, a significant fraction of the incoming energy is reflected back to the ocean, thus modifying the wave spectral content in the up-wave area. Since reflected wave components are linearly obtained from incident wave components, the cross-spectrum between incident and reflected wave fields is non-zero, resulting in the non-homogeneous spectrum $S_{tot}$ formulated in \eqref{eq:spectra}. For an infinite number of rows and using the ODC strategy, $S_{tot}(x,f)$ is illustrated in Fig. \ref{fig:Stot}, where $x = 0$ represents the position of the first row and $x<0$ is the up-wave zone. Marked interaction patterns are observed, especially around Bragg frequencies where reflection is more significant. At each frequency, the longitudinal positions of consecutive spectral energy peaks (i.e. anti-nodes, where interaction is constructive) are separated by half the wavelength. Such spectral interaction patterns are the equivalent, for a polychromatic, random wave field, of the standing wave patterns which result from the reflection of monochromatic waves. At the Bragg frequencies (where the reflection magnitude is 1), the energy content oscillates along the $x$ position, taking values between $0$ (at reflection nodes, where interaction is destructive) and $4 S_0$ (at anti-nodes).

\begin{figure}
\begin{center}
\includegraphics[width= 1\linewidth]{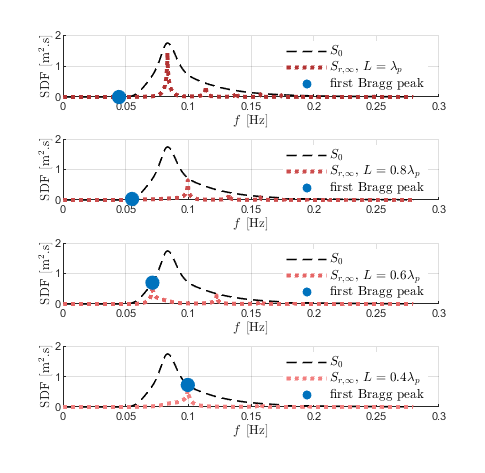}
  \caption{Sensitivity of Bragg peak locations to row-to-row spacing. Infinite array under the ODC strategy.}
\label{fig:Lsens}
\end{center}
\end{figure}

The localised amplifications in spectral energy content discussed above are certainly to be taken into account when designing the OWSC array. Since the  reflection peak locations, in the frequency space, are dictated by the Bragg resonance condition $2L = n\lambda(f^*_n)$, changing the row-to-row distance could shift those peaks to frequencies where wave energy is weaker. In that perspective, for an array with a number of rows $N\rightarrow \infty$, under the ODC strategy, Fig. \ref{fig:Lsens} shows the reflected wave spectra $S_r$ obtained for several values of $L$ (assuming that the WSA remains valid for those values smaller than $\lambda_p$, as discussed in the introduction). In each graph, the first Bragg peak is highlighted by means of a circular, blue marker. Setting $L$ to an appropriate value can prevent the spectral peak from coinciding with a Bragg reflection peak, either by ensuring that the spectral peak remains between the first and second Bragg peaks ($L = 0.8\lambda_p$,  $L = 0.6\lambda_p$ in Fig. \ref{fig:Lsens}) or by setting $L$ to a value small enough so that the first Bragg peak is above the spectral peak ($L = 0.4\lambda_p$ in Fig. \ref{fig:Lsens}).

\section{Discussion and conclusion}
\label{s:conclusion}

This preliminary study explores the concept of an artificial canopy, designed to reflect and absorb wave energy. Rows of relatively small OWSCs are the elementary ``building blocks'' of the canopy. Individual row characteristics, together with the row-to-row spacing, determine the array properties, seen as a meta-material. In that respect, the results of Section \ref{s:singlerow} explore how parameter tuning of individual rows (seen as ``classical'' wave energy converters) relates to its reflection-transmission-absorption properties (which govern array properties at the macro scale). 

Array properties are derived from OWSC properties and row-to-row spacing, through a recursive method based on the WSA. In spite of its relative simplicity, the proposed framework lends itself to a wealth of parametric analysis and optimisation, of which the present work only covers a handful of examples. Indeed, the case study of Sections \ref{s:casestudy} and \ref{s:results} is limited to regularly-spaced, identical rows, while interesting results could also be obtained by allowing inhomogeneous properties across the array. However it may be argued that, from an industry-oriented and cost-reduction perspective, considering identical devices is a sensible starting point.

Results of Section \ref{s:results} suggest that interaction effects, within arrays of small-sized, fixed structures, are unlikely to provide effective coastal protection: the resulting band-gap frequency intervals, where reflection is enhanced, are too narrow to cover a significant fraction of the incoming wave spectrum. Note, however, that such conclusions may change if viscous losses at each row were taken into account. 

In contrast, arrays of wave-absorbing, moving structures are able to capture almost all the incident wave energy, except around narrow Bragg reflection peaks. The behaviour of such arrays, even with a relatively small number of rows (e.g. $N = 5$), is well described by the ``infinite-number-of-rows'' approximation, for which the incident energy is entirely shared between absorption and reflection. At Bragg peaks, reflection tends to unity at the expense of power absorption, but in any case, wave energy is effectively prevented from reaching the shoreline.  

Regardless of how it is achieved, significant wave reflection results in interaction patterns in the ``up-wave'' area, whence spectral energy content can be greatly enhanced in periodically-spaced longitudinal positions. Those possibly undesirable effects may be mitigated by tuning the row-to-row spacing parameter, which governs the Bragg peak frequencies.

The two control strategies explored were primarily chosen because they allow the explicit derivation of $\hat{T}$ and $\hat{R}$, without resorting to the specifics of any particular mode of motion, device material or PTO system. Comparing their results suggests that there is room for joint optimisation of the array size, layout and control tuning parameters. 

Nevertheless, all conclusions above need to be nuanced and enriched, by considering more refined modelling assumptions. At the scale of individual rows, it will be interesting to investigate how specific modes of motion (rigid or flexible), and material choices, influence precisely the position of $\hat{T}$ and $\hat{R}$ for freely-moving devices (see Fig. \ref{fig:r_t_moving}), and how those choices may affect the PTO and control system requirements, needed to achieve target transmission-reflection-absorption properties using control. In the same perspective, the simplistic representation of CCC and ODC strategies must be detailed further, with a view to the difficulties which would arise from their practical implementation, such as the non-causality of the corresponding control law, or the need for large amounts of reactive power \cite{Ringwood_2014}. Finally, loss models will be included, both at the fluid interface (in the form of non-linear viscous drag) and internally (in the form of PTO losses), with possible effects on Bragg resonance.

At the canopy scale, the distribution of power absorption and mechanical loads across the array should be further examined. In particular, over-utilised front rows and under-utilised back rows would certainly be sub-optimal in an industrial production perspective. The effect of sea bottom topography could also deserve investigation. For example, a sloping bottom could be considered, in the spirit of \cite{Sabuncu_1992}. Finally, a more realistic representation of incoming wave directionality would enhance the practical value of the proposed modelling framework.

\ifCLASSOPTIONcaptionsoff
  \newpage
\fi

\bibliographystyle{IEEEtran}


\end{document}